\documentclass[preprint,prd,aps,showpacs,showkeys,nofootinbib]{revtex4-1}
\usepackage{amsmath}
\usepackage{amssymb}
\usepackage{amsfonts}
\usepackage{amssymb}
\usepackage{dcolumn}
\usepackage{bm}
\usepackage{graphicx}
\usepackage{xcolor}
\usepackage{array}
\usepackage{verbatim}
\usepackage{hyperref}
\usepackage{orcidlink}
\usepackage{ulem}

\begin{document}
	\title{ Polynomial Potential Inflation in the ACT Era: From CMB to Primordial Black Holes}
	\author{Zhi-Zhang Peng \orcidlink{0000-0001-9857-5504} $^{1}$}
	
	\author{Zu-Cheng Chen \orcidlink{0000-0001-7016-9934} $^{2,3}$}
	\email{zuchengchen@hunnu.edu.cn}
	
	\author{Lang Liu \orcidlink{0000-0002-0297-9633} $^{4}$}
	\email{liulang@bnu.edu.cn}

	\affiliation{$^1$School of Physics and Astronomy, Beijing Normal University, Beijing 100875, People’s Republic of China}
	
	\affiliation{$^2$Department of Physics and Synergetic Innovation Center for Quantum Effects and Applications, Hunan Normal University, Changsha, Hunan 410081, China}
	
	\affiliation{$^3$Institute of Interdisciplinary Studies, Hunan Normal University, Changsha, Hunan 410081, China}
	
	\affiliation{$^4$Faculty of Arts and Sciences, Beijing Normal University, Zhuhai 519087, China}


\begin{abstract}
The recent measurements from the Atacama Cosmology Telescope (ACT) favor a higher value of the scalar spectral index $n_s$ compared to the Planck data, challenging many well-established inflationary models. In this work, we investigate the viability of polynomial potential inflation in light of the latest ACT data, systematically analyzing cases from $n=2$ to $n=5$. By exploring the parameter space and deriving constraints on the model coefficients, we find that the cubic to quintic models can provide a good fit to the data, while the quadratic model struggles to simultaneously accommodate the ACT data and the requirement of sufficient inflation. Notably, the quintic case ($n=5$) not only matches cosmic microwave background (CMB) observations but also produces an inflection point that simultaneously triggers primordial black hole formation and generates a scalar-induced gravitational wave. These findings establish higher-order polynomial potentials as compelling frameworks and reconcile precision CMB measurements with multi-messenger probes of early-universe physics.
\end{abstract}
\maketitle

\section{Introduction} 	
Inflation, a period of accelerated expansion in the early Universe, has become an important paradigm in modern cosmology~\cite{Guth:1980zm, Linde:1981mu, Albrecht:1982wi}. It provides a compelling solution to several longstanding puzzles in the standard Big Bang model, such as the flatness, horizon, and monopole problems. Moreover, inflation sets the initial conditions for the formation of large-scale structures by generating a nearly scale-invariant spectrum of primordial perturbations~\cite{Mukhanov:1981xt, Hawking:1982cz, Starobinsky:1982ee, Guth:1982ec}. The inflationary framework has been extensively tested and constrained using various cosmological observations, particularly the precise measurements of the cosmic microwave background (CMB) anisotropies.

The latest data release from the Atacama Cosmology Telescope (ACT) has provided new insights into the early Universe and the nature of inflation~\cite{ACT:2025fju, ACT:2025tim}. The joint analysis of ACT and Planck data yields a higher value for the scalar spectral index, $n_s = 0.9709 \pm 0.0038$, compared to the value reported by Planck alone~\cite{Planck:2018vyg}. When combined with additional data from CMB experiments, baryon acoustic oscillations (BAO), and the Dark Energy Spectroscopic Instrument (DESI), the value of $n_s$ further increases to $0.9743 \pm 0.0034$~\cite{ACT:2025fju, ACT:2025tim}. This new observational result has significant implications for inflationary models. Some well-established models, such as the Starobinsky model~\cite{Starobinsky:1980te}, which were previously consistent with the data, now lie on the $2\sigma$ boundary and are disfavored. The shift in the $r-n_s$ plane, where $r$ is the tensor-to-scalar ratio, highlights the need to revisit and refine inflationary models in light of the updated observational constraints~\cite{Kallosh:2025rni,Gialamas:2025kef,Kim:2025dyi, Frob:2025sfq, Berera:2025vsu, Dioguardi:2025mpp, Aoki:2025wld,Dioguardi:2025vci,Salvio:2025izr,Brahma:2025dio,Gao:2025onc,Drees:2025ngb,He:2025bli,Zharov:2025evb,Yin:2025rrs,Liu:2025qca,Gialamas:2025ofz,Maity:2025czp,Yi:2025dms,Addazi:2025qra,Byrnes:2025kit,McDonald:2025odl,Yogesh:2025wak,Haque:2025uis,Haque:2025uri}.

In this paper, we propose a toy model with a polynomial potential to explain the recent ACT data. This model is characterized by a polynomial form of the inflaton potential, $V(\phi)=V_0\left(1+ \sum_{i=1}^{n} a_i \phi^i\right)$, where $\phi$ is the inflaton  and $a_i$ are the polynomial coefficients. The polynomial form here is configured as a hill-top type~\cite{Boubekeur:2005zm}, which better aligns with the observational trend of a small tensor-to-scalar ratio $r$. This form finds a natural motivation in supergravity model building~\cite{Hodges:1989dw,Cirigliano:2004yh,Lyth:2006ec,Allahverdi:2006wt,Kallosh:2010xz, Nakayama:2013jka, Nakayama:2014wpa,Linde:2014nna, Li:2014zfa,Ferrara:2016fwe,Musoke:2017frr, Kallosh:2017wku, Wolf:2024lbf,Bernal:2021qrl}, where it can be viewed as a Taylor expansion of an effective potential, with the coefficients $a_i$ linked to fundamental parameters in the Kähler potential and superpotential.  By adjusting the coefficients and the degree of the polynomial, one can obtain a wide range of inflationary dynamics and predictions for cosmological observables. In particular, the hill-top configuration ($a_2<0$) arises naturally in such high-energy physics contexts, providing a well-motivated framework that connects inflationary cosmology to fundamental theory. We examine the viability of this model by confronting its predictions with the updated observational constraints on $n_s$ and $r$. 

We find that the quadratic case ($n=2$) is in tension with the higher $n_s$ value favored by ACT data, unless one assumes an unusually large e-folding number. In contrast, the cubic ($n=3$) and quartic ($n=4$) cases provide sufficient parametric freedom to accommodate the ACT constraints. The additional parameter $a_5$ enables better tuning of the spectral index $n_s$ on CMB scales, while also allowing the potential to develop an inflection point for $n\ge3$ on  large or small scales~\cite{Linde:2007jn,Itzhaki:2007nk,Badziak:2008gv,Enqvist:2010vd,Hotchkiss:2011am,Gao:2015yha,Bernal:2021qrl,Di:2017ndc,Cheng:2018qof}. Such a feature can trigger a phase of ultra-slow-roll inflation, significantly amplifying curvature perturbations on small scales and potentially leading to the production of primordial black holes (PBHs)~\cite{Garcia-Bellido:2017mdw,Germani:2017bcs,Ballesteros:2017fsr}. However, for $n\le4$, achieving the required power spectrum enhancement ($\mathcal{P}_{\mathcal{R}} \sim 0.01$) typically conflicts with the CMB constraint on $n_s$~\cite{Ballesteros:2017fsr,Ballesteros:2020qam}, a tension further heightened by the ACT results. The $n=5$ case is notable in that it can simultaneously satisfy CMB bounds and produce a sufficiently sharp enhancement in perturbations to facilitate PBH formation. Moreover, the latter is accompanied by the generation of scalar-induced gravitational waves (SIGWs), offering a complementary observational signature.

The remainder of this paper is organized as follows. In Sec.~\ref{II}, we review the formalism for slow-roll inflation and analyze the CMB predictions for the polynomial inflation model with $n=2$, $3$ and $4$, contrasting them with the ACT data. The core results for the $n=5$ case are presented in Sec. \ref{III}, including its fit to the ACT data, the mechanism for PBH formation, and the predicted SIGW background. We discuss the implications of our findings and conclude in Sec. \ref{IV}. In what follows, we work in units where $c=\hbar=1$ and $M_{\mathrm {pl}}^2=1/8\pi G=1$.

\section{Polynomial potential inflation and ACT results}\label{II}
We consider a general polynomial potential for the inflaton field $\phi$ of the form
\begin{align}\label{vphigen}
V(\phi)=V_0\left(1+ \sum_{i=1}^{n} a_i \phi^i\right),
\end{align}
where $V_0$ sets the energy scale of inflation, $a_i$ are dimensionless coefficients. We focus on the cases $n=2$, $n=3$ and $n=4$. It is  conventional to define the  slow-roll parameters $\epsilon$ and $\eta$ as
\begin{align}\label{srpara}
\epsilon = \frac{1}{2}\left(\frac{V_\phi}{V}\right)^2, \quad \eta =  \frac{V_{\phi\phi}}{V}.
\end{align}
Up to the first order slow-roll approximation, the power spectrum of the curvature perturbation, its spectral index and the tensor-to-scalar ratio are expressed as
\begin{align}\label{nsr}
   \mathcal{P}_{\mathcal{R}}&=\frac{V}{24 \pi^2 \epsilon}\,, \nonumber \\ 
    n_s &= 1 - 6 \epsilon +2 \eta\,,   \nonumber\\
    \quad r &=16 \epsilon\,.
\end{align}
Next, we will use these basic slow-roll formulae to confront the predictions of polynomial inflation with the ACT data.

\subsection{$\mathbf{n=2}$} 
\begin{figure*}[htpb]
    \centering
    \includegraphics[width=0.49\linewidth]{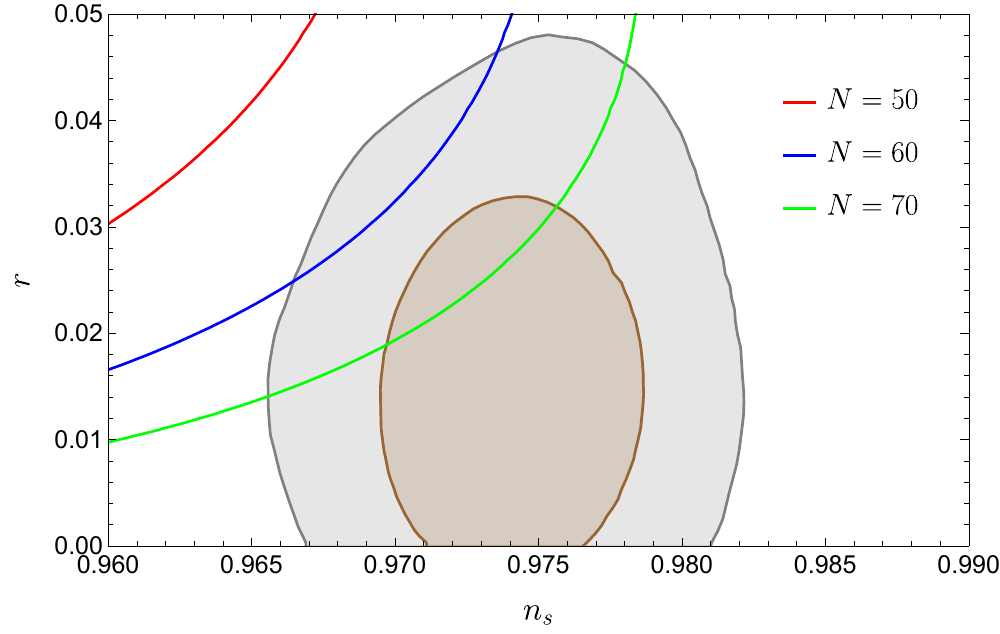}
    \includegraphics[width=0.49\linewidth]{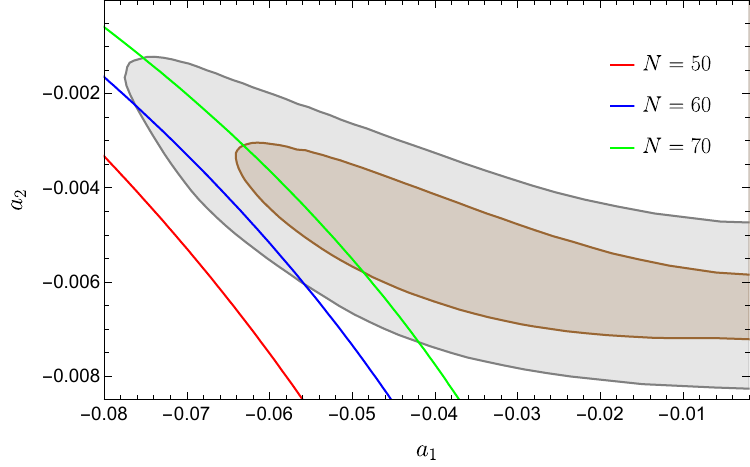}
    \caption{(Left) Scalar spectral index $n_s$ and tensor-to-scalar ratio $r$ predicted by the quadratic potential for $N=50$, $N=60$ and $N=70$, compared with observational constraints. The contours are the $1\sigma$ and $2\sigma$ observational constraints from the  P-ACT-LB data  combined with B-mode measurements from the BICEP and Keck telescopes at the South Pole (BK18), referred to as P-ACT-LB-BK18~\cite{ACT:2025fju, ACT:2025tim}. (Right) Corresponding parameter values of $a_1$ and $a_2$ consistent with observational bounds. The brown and gray color regions are the constraints from $1\sigma$ and $2\sigma$ observational constraints, respectively.}
    \label{quadratic_nsr}
\end{figure*}

For $n=2$, we consider the following polynomial potential
\begin{align}\label{vphi2}
    V(\phi)=V_0(1+a_1 \phi+a_2 \phi^2)\,.
\end{align}
Combining Eqs. \eqref{srpara}-\eqref{vphi2}, the dimensionless parameters $a_1$ and $a_2$ can be analytically expressed by $n_s$ and $r$. Note that the expression actually includes $\phi_*$, which corresponds to the pivot scale exiting horizon. However, the shift of the inflaton does not alter the form of the inflaton potential. As a result, the only change is that the dimensionless parameters undergo a rescaling. Therefore, we can arbitrarily choose the value of $\phi_*$. Without loss of generality, we set $\phi_*=0$ thereafter. We obtain  the following expressions which are determined by $n_s$ and $r$ 
\begin{align}\label{a1a2}
    a_1=-\frac{\sqrt{r}}{2\sqrt{2}}\,,\quad a_2=\frac{1}{32}(-8+8n_s+3r)\,.
\end{align}
{The magnitude $|a_1| = \sqrt{r}/(2\sqrt{2})$ follows directly from $r = 16\epsilon = 8a_1^2$. The negative sign is then fixed by requiring $V_\phi(\phi_*) < 0$, ensuring that the inflaton rolls downhill in the positive field direction consistent with our hill-top configuration.}
 For hill-top inflation, the values of scalar field at the end of inflation $\phi_e$ is determined by the condition
\begin{align}
   | \eta(\phi_e)|=1\,. 
\end{align}
{This condition marks the breakdown of the slow-roll approximation and the end of the inflationary phase. It should not be confused with the true vacuum state ($V_\phi = 0$), which the inflaton reaches only later during the reheating epoch. For hill-top models, $|\eta| = 1$ is typically reached before $\epsilon = 1$ because the potential curvature dominates over the slope near the maximum \cite{Liddle:2000cg,Baumann:2009ds,Boubekeur:2005zm}.}
We then get
\begin{align}\label{phie}
\phi_e=\frac{4\sqrt{2}\sqrt{r} - \sqrt{2}\sqrt{64 - 64n_s^2 + 16r - 48n_s r - 9r^2}}{-8 + 8n_s + 3r}\,.
\end{align}
The $e$-folding number $N$ of canonical single-field slow-roll inflation is then evaluated as
\begin{align}
    N=\int^{\phi_e}_{\phi_*}\frac{\mathrm{d}\phi}{\sqrt{2\epsilon}}\,.
\end{align}
Finally, we can derive the analytical expression for $N$
\begin{align}
    N=-\frac{\left(a_1^2-4 a_2\right) (\log (a_1)-\log (a_1+2 a_2 \phi_e))+2 a_2 \phi_e (a_1+a_2 \phi_e)}{8 a_2^2}\,,
\end{align}
where $a_1$, $a_2$ and $\phi_e$ are given by Eq.~\eqref{a1a2} and Eq.~\eqref{phie}. Next, we can impose constraints on the dimensionless parameters $a_1$ and $a_2$ using ACT data. On the other hand, $N$ is uniquely determined by $n_s$ and $r$. By allowing the $N$ to vary within the range $50\sim60$, the model's predictions for $n_s$ and $r$ are derived, leading to further constraints on the parameter space.  The results are shown in Fig.~\ref{quadratic_nsr}. The left panel illustrates the theoretical predictions for the scalar spectral index $n_s$ and the tensor-to-scalar ratio $r$ when fixing the $e$-folding number $N$, whereas the right panel depicts the parameter space of the dimensionless parameters $a_1$ and $a_2$. The brown and gray regions denote the $1\sigma$ ($68\%$ C.L.) and $2\sigma$ ($95\%$ C.L.) confidence regions, respectively, are superimposed from the Planck and ACT datasets. To ensure consistency with with the $1\sigma$ constraint of
the observational data, the dimensionless parameters $a_1$ and $a_2$ should satisfy
\begin{align}
    -0.06<a_1<0\,,\qquad -0.007<a_2<-0.003\,.
\end{align}
When accounting for the $e$-folding number $N$, we observe that the dimensionless parameters corresponding to $N=60$ lies outside the $1\sigma$ confidence region derived from the ACT data. When $N=50$, the corresponding parameter space does not even fall within the $2\sigma$ region. To reconcile with the observational constraints, larger values of $N$ (e.g., $N=70$) are required.

\subsection{$\mathbf{n=3}$} 
For $n=3$, we consider the following polynomial potential
\begin{align}\label{vphi3}
    V(\phi)=V_0 (1+a_1 \phi+a_2 \phi^2+a_3 \phi^3)\,.
\end{align}
The inclusion of an additional parameter renders the scalar spectral index $n_s$
and tensor-to-scalar ratio $r$  insufficient to impose stringent constraints on the full parameter space. By introducing the running of the spectral index $\alpha_s$ (defined as $\mathrm{d}n_s/\mathrm{d}\ln k $), we could derive a viable parameter region consistent with ACT data. To second order, the following two equations hold
\begin{align}\label{dotslow}
    \frac{\mathrm{d}\epsilon}{\mathrm{d}\ln k}&=2\epsilon(2\epsilon-\eta)\,,\\
    \frac{\mathrm{d}\eta}{\mathrm{d}\ln k}&=2\epsilon\eta-\xi\,,
\end{align}\
with
\begin{align}\label{xi}
    \xi=\frac{V_\phi V_{\phi\phi\phi}}{V^2}\,.
\end{align}
{Here $k$ denotes the comoving wavenumber of primordial perturbations. The scalar spectral index $n_s$ characterizes the scale dependence of the primordial power spectrum, $\mathcal{P}_s(k) \propto k^{n_s-1}$, arising because different $k$-modes exit the Hubble horizon at different epochs during inflation and thus probe slightly different values of the slow-roll parameters. Observational constraints are quoted at the pivot scale $k_* = 0.05~\mathrm{Mpc}^{-1}$.}
Note that $\xi=0$ for the case of $n=2$. In terms of Eq.~\eqref{nsr} and Eqs.~\eqref{dotslow}-\eqref{xi}, The running of the spectral index $\alpha_s$ can be written as
\begin{align}\label{run}
    \alpha_s=16\epsilon\eta-24\epsilon^2-2\xi\,.
\end{align}
Combining Eqs.~\eqref{srpara}, \eqref{vphi3} and \eqref{run}, and setting $\phi_*=0$, dimensionless parameters are given by 
\begin{align}
    a_1=-\frac{\sqrt{r}}{2\sqrt{2}}\,,\quad a_2=\frac{1}{32}(-8+8n_s+3r)\,,\quad a_3=\frac{32\sqrt{2}\alpha_s+16\sqrt{2}r-16\sqrt{2}n_s r-3\sqrt{2}r^2}{192\sqrt{r}}\,.
\end{align}
The computation of $\phi_e$ and $N$ follows the methodology outlined in the preceding subsection. Their explicit analytical forms are omitted here due to their cumbersome expressions.

The cubic potential can be simplified by a field shift $\phi=\phi+c$ to eliminate the linear term, yielding $V(\phi)=\tilde{V}_0(1+\tilde{a}_2\phi^2+\tilde{a}_3\phi^3)$without loss of generality. Physical observables are invariant under such a translation. In the slow-roll approximation, the hill-top-type potential (with $\tilde{a}_2<0$) yields the scaling relations $n_s\approx1-2/N\,, \alpha_s\approx-2/N^2$. For $N=50-60$, one obtains $\alpha_s\sim-0.0006$ to $-0.0008$. The presence of the cubic term slightly modifies this scaling.

 In the present paper, we set $\alpha_s=-0.001$, which is consistent with the joint constraints from Planck, BAO and BK15, $\alpha_s=-0.006 \pm 0.013$ at $k=0.002 \mathrm{Mpc}^{-1}$ ($95\% $ C.L.)~\cite{Planck:2018vyg}.  
 For the case $ n = 3 $, the values of $ a_1 $ and $ a_2 $ coincide with those for $ n = 2 $, a result arising from the choice $ \phi_* = 0 $. In the general scenario, these parameters are functions of $ n_s $, $ r $, and $ \alpha_s $ . Our focus lies in assessing whether such a model can plausibly reconcile with the observational constraints from the ACT, rather than precisely delineating the parameter space. While different $ \phi_* $ correspond to distinct parameter spaces, the question of whether the parameter space falls within the  confidence region under the $ e $-folding number $ N $ constraints remains independent of $ \phi_* $.  

\begin{figure*}[htpb]
    \centering
    \includegraphics[width=0.7\linewidth]{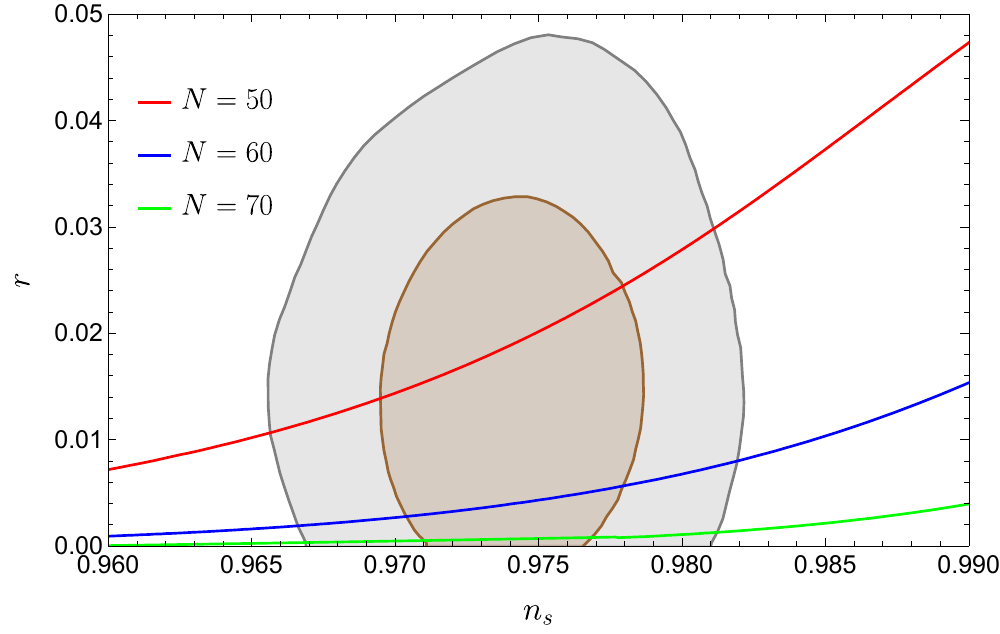}
    \caption{  The $n_s-r$ trajectories for the cubic potential, evaluated at $N=50$ (red), $60$ (blue), and $70$ (green).  The brown and gray color regions are the constraints from $1\sigma$ and $2\sigma$ observational constraints, respectively.}\label{cubic_nsr}
\end{figure*}

As illustrated in Fig.~\ref{cubic_nsr}, for $ \alpha_s = -0.001 $, the model gives a broad viable range of $ N $. Fixing $ N = 50 $, $ 60 $, and $ 70 $, respectively, the corresponding $ n_s $-$ r $ curves all intersect the $ 1\sigma $ region. The representative  dimensionless parameter values and corresponding predictions are shown in Tab.~\ref{cubic_nsr_tab}. This demonstrates that the $ n = 3 $ case retains sufficient parameter space to accommodate ACT data.

\begin{table}
\centering
		\begin{tabular}{>{\centering}p{1.5cm}>{\centering}p{1.5cm}>{\centering}p{1.5cm}>{\centering}p{1.5cm}>{\centering}p{1.5cm}>{\centering}p{1.5cm}>{\centering}p{1.5cm}}
			\hline
			\hline
			$\textit{Set}$ & $a_1$ & $a_2$ & $a_3$  & $n_s$ & $r$ & $N$ \tabularnewline
			\hline
			
			$1$ & $-0.0493$ & $-0.0045$ & $-0.0013$  & $0.9745$ & $0.0195$ & $50$\tabularnewline
			
			$2$ & $-0.0212$ & $-0.0064$ & $-0.0037$  & $0.9732$ & $0.0036$ & $60$ \tabularnewline

            $3$ & $-0.0101$ & $-0.0061$ & $-0.0082$  & $0.9753$ & $0.0008$ & $70$  \tabularnewline
			\hline
			\hline
		\end{tabular}
		\caption{Representative points and corresponding predictions for cubic polynomial inflation.}\label{cubic_nsr_tab}
	\end{table}
    
\subsection{$\mathbf{n=4}$} 
For $n=4$, The potential is
\begin{align}
  V(\phi)=V_0 (1+a_1 \phi+a_2 \phi^2+a_3 \phi^3+a_4 \phi^4)\,.  
\end{align}
The slow-roll parameters are given by
\begin{align}
    \epsilon=\frac{\left(a_1+2 a_2 \phi +3 a_3 \phi ^2+4 a_4 \phi ^3\right)^2}{2 \left(1+a_1\phi+a_2\phi ^2+a_3 \phi ^3+a_4\phi ^4\right)^2}\,,\quad \eta= \frac{2 a_2+6 a_3 \phi +12 a_4\phi ^2}{1+a_1\phi+a_2 \phi ^2+a_3 \phi ^3+a_4 \phi ^4}\,.
\end{align}
 We focus on the parameter region where $a_3>0$ and $a_4<0$, which yields a potential with a flat plateau suitable for sustained inflation and a red-tilted scalar spectrum. To compare with high-precision cosmological data, it is essential to consider not only the scalar spectral index $n_s$ and the tensor-to-scalar ratio $r$, but also the running $\alpha_s$ and the running of the running $\beta_s$ of the spectral index. The latter is given by
\begin{align}
    \beta_s=-192\epsilon^3 + 192\epsilon^2\eta - 32\epsilon\eta^2 - 24\epsilon\xi + 2\eta\xi + 2\omega\,,
\end{align}
with
\begin{align}
    \omega=\frac{V_\phi^2V_{\phi\phi\phi\phi}}{V^3}\,.
\end{align}
Note that $\omega=0$ for $n=3$, but becomes non-zero for $n\ge4$, providing an additional degree of freedom to fit the data. Expressing the coefficients $a_i$ in terms of the observables $(n_s, r, \alpha_s,\beta_s)$ evaluated at the pivot scale $\phi_*=0$ yields

\begin{align}
     a_1&=-\frac{\sqrt{r}}{2\sqrt{2}}\nonumber \\
      a_2&=\frac{1}{32}(-8+8n_s+3r)     \nonumber \\
      a_3&=\frac{32\sqrt{2}\alpha_s+16\sqrt{2}r-16\sqrt{2}n_s r-3\sqrt{2}r^2}{192\sqrt{r}}\nonumber \\
      a_4&=\frac{512\beta_s - 256\alpha_s + 256\alpha_s\,n_s + 128\,r - 288\alpha_s\,r - 256\,n_s\,r}{3072\,r} \nonumber \\
      &+\frac{  128\,n_s^2\,r - 120\,r^2 + 120\,n_s\,r^2 + 15\,r^3}{3072\,r}
\end{align}

These relations allow us to map observational constraints directly onto the parameter space of the model. For the quartic potential, the cubic equation $V_\phi=0$ determines the critical points. {The critical points of the potential, determined by $V_\phi = 0$, characterize the structure of the scalar potential (its maxima and minima). This is distinct from the end-of-inflation condition $|\eta(\phi_e)| = 1$, which identifies where slow-roll violation terminates the accelerated expansion.} Assuming one real root $\gamma$ and a pair of complex conjugate roots $ \alpha \pm i \beta$, the integral can be performed analytically via partial fraction decomposition, giving
\begin{align}
N = \left[ \frac{1}{8a_{4}} \phi^{2} + \frac{a_{3}}{16a_{4}^{2}} \phi + A \ln|\phi - r| + \frac{B}{2} \ln\left( (\phi - \alpha)^{2} + \beta^{2} \right) + \frac{B\alpha + C}{\beta} \arctan\left( \frac{\phi - \alpha}{\beta} \right) \right] \Bigg|_{\phi_{*}}^{\phi_{e}}
\end{align}
where the coefficients $A$, $B$, and $C$, determined by partial fraction decomposition, are real numbers that depend on $a_1$, $a_2$, $a_3$, $a_4$, $\gamma$, $\alpha$, and $\beta$. The end point $\phi_e$ is obtained from $|\eta(\phi_e)|=1$ and also depends on the parameters $a_i$.

\begin{figure*}[htpb]
    \centering
     \includegraphics[width=0.7\linewidth]{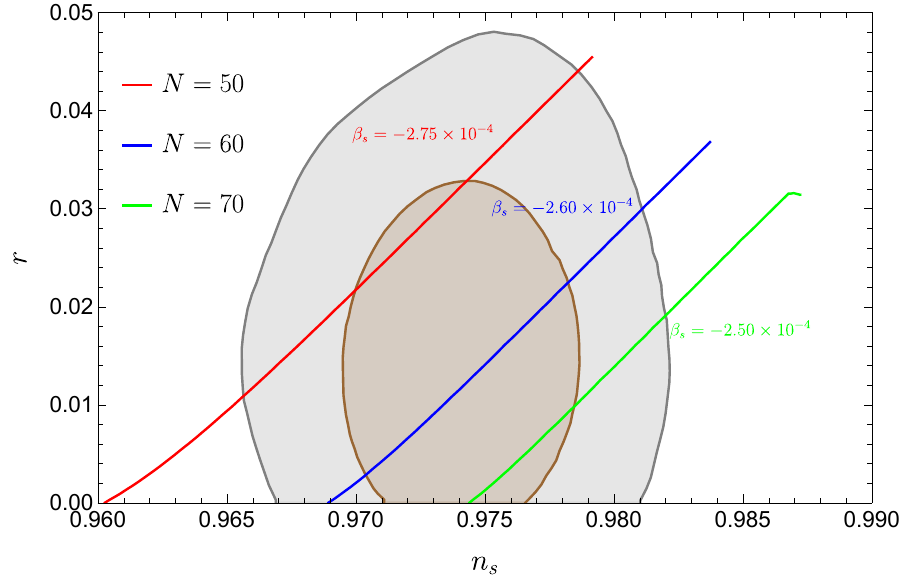}
    \caption{ The $n_s-r$ trajectories for the quartic potential, evaluated at $N=50$ (red), $60$ (blue), and $70$ (green). Each curve corresponds to a specific value of the running of the running: $\beta_s=-2.75\times10^{-4}$, $-2.60\times10^{-4}$, $-2.50\times10^{-4}$, respectively.  The brown and gray color regions are the constraints from $1\sigma$ and $2\sigma$ observational constraints, respectively.}\label{quartic_nsr}
\end{figure*}

We conducted a comprehensive parameter scan and found that $\alpha_s> 0$ and $\beta_s<0 $ facilitate better parameter tuning.  We fix the running of the spectral index as $\alpha_s=0.006$  and explore the dependence on $\beta_s$. The right panel of Fig.~\ref{quartic_nsr} shows the model predictions for the $n_s-r$ relation at $ N = 50 $, $ 60 $ and $ 70 $, with representative points marking the $1\sigma$ confidence region shown in Tab.~\ref{quartic_nsr_tab}. The black and brown shaded areas denote the $1\sigma$ and $2\sigma$ observational bounds, respectively, from a combination of Planck and BAO data.

The key outcome is that, for a fixed $\alpha_s$, a larger e-folding number $N$ requires a larger (less negative) value of $\beta_s$. This trend can be understood physically: a larger $\beta_s$ corresponds to a larger absolute value of the coefficient $|a_4|$, which flattens the potential near the maximum and prolongs the inflation. Consequently, the predicted curves in the $n_s-r$ plane shift toward higher $n_s$  as $|a_4|$ increases, improving the consistency with the latest ACT and Planck measurements.

\begin{table}
\centering
		\begin{tabular}{>{\centering}p{1.5cm}>{\centering}p{1.5cm}>{\centering}p{1.5cm}>{\centering}p{1.5cm}>{\centering}p{1.5cm}>{\centering}p{1.5cm}>{\centering}p{1.5cm}>{\centering}p{1.5cm}}
			\hline
			\hline
			$\textit{Set}$ & $a_1$ & $a_2$ & $a_3$ & $a_4$ & $n_s$ & $r$ & $N$ \tabularnewline
			\hline
			
			$1$ & $-0.0582$ & $-0.0044$ & $-0.0090$ & $-0.1704$ & $0.9721$ & $0.0271$ & $50$\tabularnewline
			
			$2$ & $-0.0441$ & $-0.0047$ & $-0.0116$ & $-0.2795$ & $0.9755$ & $0.0156$ & $60$ \tabularnewline

            $3$ & $-0.0226$ & $-0.0056$ & $-0.0223$ & $-1.0197$ & $0.9761$ & $0.0041$ & $70$  \tabularnewline
			\hline
			\hline
		\end{tabular}
		\caption{Representative points and corresponding predictions for quartic polynomial inflation.}
		\label{quartic_nsr_tab}
	\end{table}

\section{Small-Scale Phenomenology with Quintic Potential}\label{III}

The analysis in the first part  demonstrated that the ACT data can be well accommodated by a polynomial potential including terms up to the quartic order. 

However, the physics of inflation on smaller scales, which are not directly probed by the CMB, may involve more complex dynamics. In particular, the formation of PBHs requires a significant enhancement of the primordial curvature perturbations on small scales, typically by a factor of about $10^7$ compared to the CMB scale. Such an enhancement is difficult to achieve with a smooth potential limited to lower order terms. Therefore, we need a higher-order potential, which can introduce an inflection point, leading to a period of ultra-slow-roll inflation that amplifies the perturbations.

To explore this possibility, we extend the polynomial potential to include a quintic term. In the following, we discuss the phenomenology of the quintic potential,including primordial curvature perturbations, PBHs and SIGWs.
\subsection{Primordial curvature perturbations}
The potential up to a quintic term is 
\begin{align}
V(\phi) = V_0(1+ a_1 \phi + a_2 \phi^2 + a_3 \phi^3 + a_4 \phi^4+a_5 \phi^5)\,.
\end{align}
{For the quintic potential ($n = 5$), we adopt a different computational scheme from the cases $n = 2$ to $n = 4$ treated in Section II. Here, the field translation $\phi \to \phi + C$ is used to set $a_2 = 0$, simplifying the subsequent analysis. Unlike the previous cases where we set $\phi_* = 0$ exactly, here the initial field value $\phi_*$ is not fixed to zero but is assumed to be small.} 
 Since the inflaton rolls down from near $\phi=0$, we can neglect terms higher than the third order when $\phi$ is in the vicinity of the origin
\begin{align}
V(\phi) \simeq V_0(1+ a_1 \phi  + a_3 \phi^3 )\,.
\end{align}
{For the analysis of large-scale observables (CMB/ACT), the field remains small, and we can safely neglect the higher-order terms $a_4\phi^4$ and $a_5\phi^5$. However, for SIGW and PBH production, the field approaches $\phi \sim \mathcal{O}(1)$ near the inflection point, and the quartic and quintic terms become important and must be retained.}
The slow-roll parameters are
\begin{align}
    \epsilon&=\frac{1}{2 }\left(\frac{V_\phi}{V}\right)^2=\frac{(a_1+3a_3\phi^2)^2}{2(1+a_1 \phi+a_3\phi^3)^2}\,, \\ \nonumber
    \eta&=\frac{V_{\phi\phi}}{V}=\frac{6 a_3 \phi}{1+a_1\phi+a_3\phi^3}\,.
\end{align}
The inflaton starts to roll fast after the time when $\eta\sim-1$, i.e., $\phi\simeq-\frac{1}{6 a_3}$. To determine the parameter $a_3$, we consider the e-folding number
\begin{align}\label{efolding}
     {N_1}=\int^{{\phi_1}}_{\phi_*}\frac{\mathrm{d}\phi}{\sqrt{2\epsilon}}=\left.\frac{1}{\sqrt{3 a_1 a_3}}\arctan \left(\sqrt{\frac{3a_3}{a_1}}\phi\right)\right|_{\phi_*} ^{{\phi_1}}\,.
\end{align}
{Here $\phi_1$ and $N_1$ denote the field value and e-folding number at the end of the {first slow-roll phase}, not the end of inflation. In this phase, which corresponds to large-scale (CMB/ACT) observations, the field remains small: $\phi_* < \phi < \phi_1 < 1$ (in Planck units). This justifies neglecting higher-order terms in the polynomial potential. At smaller scales, when $\phi > \phi_1$ and the field approaches the inflection point, higher-order terms become significant and must be retained---particularly for analyzing enhanced perturbations relevant to PBH formation.}
Introducing a new quantity $N_{\mathrm {tot}}$ which characterize the total number of e-folding during inflation~\cite{Baumann:2007ah,Linde:2007jn}
\begin{align}\label{etotal}
    N_{\mathrm {tot}}=\int^{\infty}_{0}\frac{\mathrm{d}\phi}{\sqrt{2\epsilon}}=\frac{\pi}{2\sqrt{3a_1a_3}}\,.
\end{align}
The scalar spectral index on large scales corresponding to the CMB can be computed using the standard slow-roll method
\begin{align}\label{ecmb}
    n_s-1\simeq2\eta(\phi_*).
\end{align}
{For $N_{\mathrm{tot}}$ to be real and positive, the product $a_1 a_3$ must be positive, requiring $a_1$ and $a_3$ to share the same sign. Meanwhile, the hill-top configuration demands $V_\phi = V_0(a_1 + 3a_3\phi^2) < 0$ during inflation, which is impossible if both $a_1$ and $a_3$ are positive. Therefore, both parameters must be negative: $a_1 < 0$ and $a_3 < 0$. This is consistent with the observed red-tilted spectrum, since $n_s - 1 \simeq 12\, a_3\, \phi_* < 0$ for $a_3 < 0$ and $\phi_* > 0$.}
Combining Eqs.~\eqref{efolding}, \eqref{etotal} and \eqref{ecmb}, we obtain
\begin{align}
    {n_s-1=\frac{2\pi}{N_{\mathrm {tot}}}\tan\left(\frac{\pi N_1}{2 N_{\mathrm {tot}}}-\arctan\left(\frac{N_{\mathrm {tot}}}{\pi}\right)\right)\,.}
\end{align}
Since $N_{\mathrm {tot}}\gg1$, we can expand the scalar spectral index as
\begin{align}\label{ns}
    n_s-1=-\frac{2\pi}{N_{\mathrm {tot}}}\cot\left(\frac{\pi {N_1}}{2 N_{\mathrm {tot}}}\right)\,.
\end{align}

To coincide with the CMB observation, we need a red-tilted power spectrum, which means $N_{\mathrm {tot}}\ge {N_1}$. If we set ${N_1}=30$  and assume  $a_3\sim\mathcal{O}(1)$,then we have
\begin{align}
    a_1\le 10^{-3} a_3^{-1}\le -10^{-3}\,.
\end{align}
On the other hand, the power spectrum of curvature perturbation is given by
\begin{align}
    \mathcal{P}_{\mathcal{R}}=\frac{V}{24 \pi^2 \epsilon}\simeq \frac{V_0}{12 \pi^2 a_1^2}\,.
\end{align}
Once $a_1$ is given, combining with $\mathcal{P}_{\mathcal{R}}=2.1\times10^{-9}$, we can immediately deduce the energy scale $V_0$.

The running of scalar spectral index $\alpha_s$ can be computed using the explicit expression given in Eq.~\eqref{ns} 
\begin{align}
    \alpha_s=\frac{dn_s}{d\ln k }=-\frac{\pi^2}{N_{\mathrm {tot}}^2}\csc^2\left(\frac{\pi {N_1}}{2 N_{\mathrm {tot}}}\right)\,.
\end{align}
The higher-order slow-roll parameters can be deduced by analogy.
To generate a significant peak in the power spectrum on small scales, we utilize the full quintic potential. We adjust the coefficients $a_4$ and $a_5$ to create an approximate inflection point at some field value, satisfying
\begin{align}
V_\phi \approx 0\,, \quad V_{\phi\phi}\approx 0\,.
\end{align}
In the vicinity of this point, the potential becomes extremely flat, causing the inflaton to enter a phase of ultra-slow-roll inflation.  {As the field evolves past the inflection point, the second slow-roll parameter $\eta$ transitions from positive to negative values. At the exact inflection point where $V_{\phi\phi} = 0$, we have $\eta = 0$. After crossing the inflection point, $\eta$ becomes increasingly negative, leading to the enhancement of primordial perturbations.} The location of the inflection point determines the comoving wavenumber $k_p$ at which the power spectrum peaks, and the flatness of the plateau controls the amplification factor.

{In the ultra-slow-roll regime near the inflection point, the field value $\phi$ is no longer small but approaches $\mathcal{O}(1)$. Consequently, the $a_4\phi^4$ and $a_5\phi^5$ terms, which were negligible for the large-scale analysis, now play a dominant role in determining the potential shape and the amplitude of enhanced perturbations.}
As a concrete example, we obtain a viable power spectrum consistent with CMB constraints on large scales and a pronounced peak on small scales with the following parameter set
\begin{align}
a_1 &= -2.42409 \times 10^{-4}, \quad a_3 = -3.20987654, \quad a_4 = 5.14403292, \\ \nonumber
\quad a_5&=-2.19727384,    \quad V_0 =1.501\times10^{-14}.
\end{align}
 For the parameter set, the numerical results on CMB scales are
\begin{align}
    n_s=0.9764\,, \quad r=4.8\times10^{-7} \,, \quad \alpha_s=-0.0095 \,, \quad \beta_s=-0.0001\,,
\end{align}
which are in agreement with the joint constraints from Planck 2018 ($68\%$ C.L.)~\cite{Planck:2018vyg} and BICEP/Keck XIII  ($95\%$ C.L.)~\cite{BICEP:2021xfz} at  the pivot scale $k_*= 0.05\mathrm{Mpc}^{-1}$
\begin{align}
   \quad r<0.036\,, \quad\alpha_s= 0.0011\pm 0.0099\,,\quad \beta_s=0.009\pm0.012\,.
\end{align}
The predicted $n_s$ and $r$ also consistent with the recent P-ACT-LB results~\cite{ACT:2025fju, ACT:2025tim}
\begin{align}
    n_s= 0.9743\pm0.0034\,,  \quad \alpha_s= 0.0062\pm 0.0052\,.
\end{align}
Notably, while the model matches the observed $n_s$ and satisfies the bound on $r$, its predicted running $\alpha_s = -0.0095$ shows a $3\sigma $ tension with the P-ACT-LB central value, though it remains consistent with Planck within $2\sigma$. This discrepancy presents a testable feature for future precision measurements.

\begin{figure*}[htpb]
    \centering
     \includegraphics[width=0.7\linewidth]{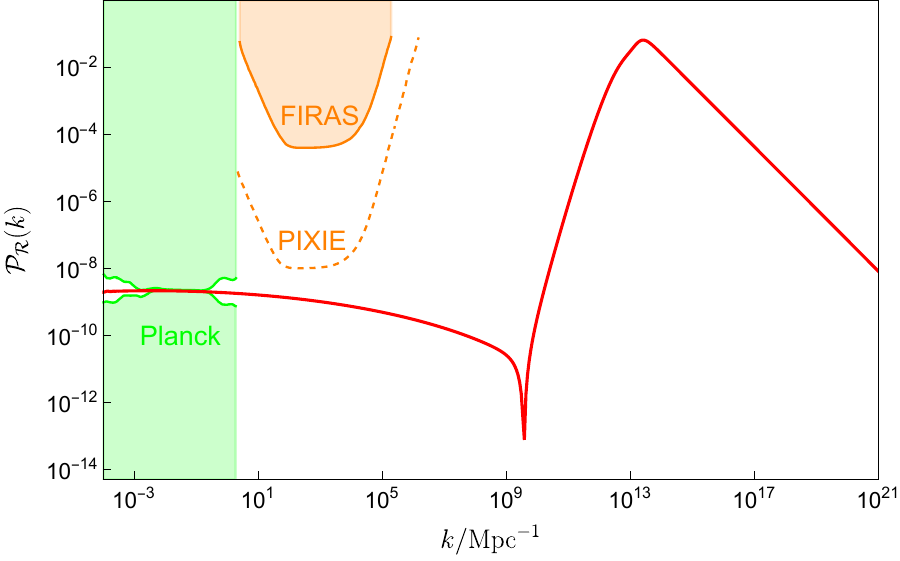}
    \caption{Power spectrum of primordial curvature perturbations.The green shaded region is excluded by CMB observations~\cite{Planck:2018jri}.
	The orange shaded region shows the current upper bound on the power spectrum from measurements of $\mu$ distortion for COBE/FIRAS~\cite{Mather:1993ij,Fixsen:1996nj}. The forecasted constraint for the distortion experiment PIXIE~\cite{Kogut:2011xw} is shown as the orange dashed line. See Ref.~\cite{Chluba:2019nxa} for the summary of constraints on the power spectrum of curvature perturbations.}\label{quintic_PRe}
\end{figure*}

On small scales, the curvature perturbations are obtained by numerically solving the Mukhanov-Sasaki equation. The result is plotted in Fig.~\ref{quintic_PRe}, which shows a prominent peak at a wavenumber of $k\approx 10^{13}{\mathrm Mpc}^{-1}$, with a maximum amplitude of approximately $\mathcal{P}_{\mathcal{R}}\approx0.065$. Such a significant enhancement would greatly facilitate the formation of PBHs.

\subsection{Primordial black holes}
PBHs are hypothesized to form in the early universe through the gravitational collapse of large‑amplitude curvature perturbations generated during inflation.  When these perturbations re-enter the Hubble horizon in the radiation-dominated era, those exceeding a critical density threshold $\delta_c $ can overcome radiation pressure and collapse into PBHs~\cite{Musco:2004ak,Musco:2008hv,Musco:2012au}. This process requires a significant enhancement of the primordial curvature power spectrum $\mathcal{P}_{\mathcal{R}}(k)$ to $\mathcal{O}(10^{-2})$ on small scales ($k \gg 1~\mathrm{Mpc}^{-1}$), contrasting with the CMB scale value of $\mathcal{O}(10^{-9})$. 

The mass of PBHs formed at horizon re-entry is related to the comoving wavenumber $k$ by
\begin{align}
M(k) = \gamma M_H \approx 30 \, M_\odot \left( \frac{\gamma}{0.2} \right) \left( \frac{g_*}{106.75} \right)^{-1/6} \left( \frac{k}{10^6 \, \mathrm{Mpc}^{-1}} \right)^{-2},
\end{align}
where $\gamma \approx 0.2$ is the collapse efficiency, $M_H$ is the horizon mass, and $g_*$ is the effective degrees of freedom. This relation links PBH formation to specific inflationary scales, enabling predictions for PBHs in observational windows such as LIGO events ($M \sim 10~M_\odot$) or dark matter ($M \sim 10^{-15}-10^{-12}~M_\odot$).

The abundance of PBHs is commonly described within the Press–Schechter formalism under the assumption of Gaussian perturbations. The fraction of the universe collapsing into PBHs of mass $M$ is given by~\cite{Blais:2002gw,Josan:2009qn}
\begin{align}
\beta(M) = \frac{1}{2} \operatorname{erfc}\left( \frac{\delta_c}{\sqrt{2 \sigma^2(M)}} \right),
\end{align}
where $\operatorname{erfc}$ is the complementary error function and $\sigma^2(M)$ is the variance of the smoothed density contrast. This variance is derived from the primordial power spectrum $\mathcal{P}_{\mathcal{R}}(k)$ through
 \begin{align}
 \sigma^2(M) = \frac{16}{81} \int_0^\infty \frac{dk}{k} \, \mathcal{P}_{\mathcal{R}}(k)  \, \left(k R\right)^4 \, W^2(kR) ,
 \end{align}
with $R \approx k_{M}^{-1}$ being the smoothing scale and $W(kR)$ the window function (e.g., Gaussian: $W(kR) = \exp(-k^2 R^2/2)$). The present‑day PBH abundance relative to dark matter is then~\cite{Sasaki:2018dmp}
\begin{align}
f_{\mathrm{PBH}}(M) = \frac{\Omega_{\mathrm{PBH}}}{\Omega_{\mathrm{DM}}} \approx 2.4 \times 10^8 \, \beta(M) \left( \frac{g_*}{106.75} \right)^{-1/4} \left( \frac{M}{M_\odot} \right)^{-1/2}.
\end{align}
\begin{figure*}[htpb]
    \centering
    \includegraphics[width=0.7\linewidth]{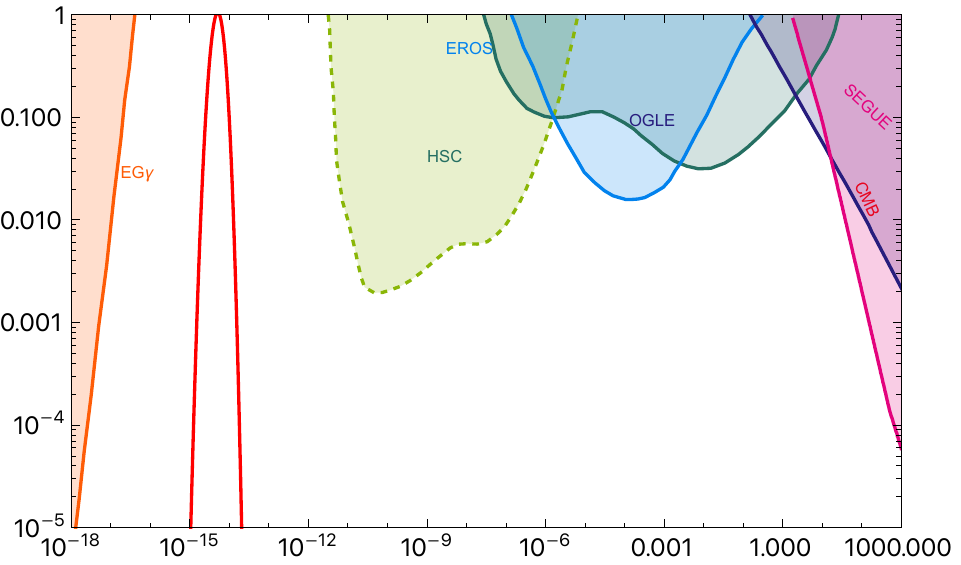}
    \caption{ Predicted PBH mass function for the parameter set given in the text.
    The shaded regions indicate current observational constraints on PBHs, derived from the extragalactic gamma-ray background (EG$\gamma$~\cite{Carr:2009jm}), the galactic 511 keV emission line ($e^{+}$~\cite{DeRocco:2019fjq,Laha:2019ssq,Dasgupta:2019cae}), gravitational lensing surveys (HSC~\cite{Niikura:2017zjd}, EROS \cite{Tisserand:2006zx}, OGLE~\cite{Niikura:2019kqi}), dynamical limits from stellar kinematics~\cite{Koushiappas:2017chw}, and CMB measurements~\cite{Poulin:2017bwe,Carr:2020gox,Green:2020jor}.}.\label{quintic_PBH}
\end{figure*}

In Fig.~\ref{quintic_PBH}, we plot the current mass spectrum of PBH produced by the power spectrum shown
 in Fig.~\ref{quintic_PRe}. One can see that PBH with a monochromatic mass around $10^{-15}$ to $10^{-14}M_\odot $  can  account for all of the dark matter.

\subsection{Scalar-induced gravitational waves}
\begin{figure*}[htpb]
    \centering
    \includegraphics[width=0.65\linewidth]{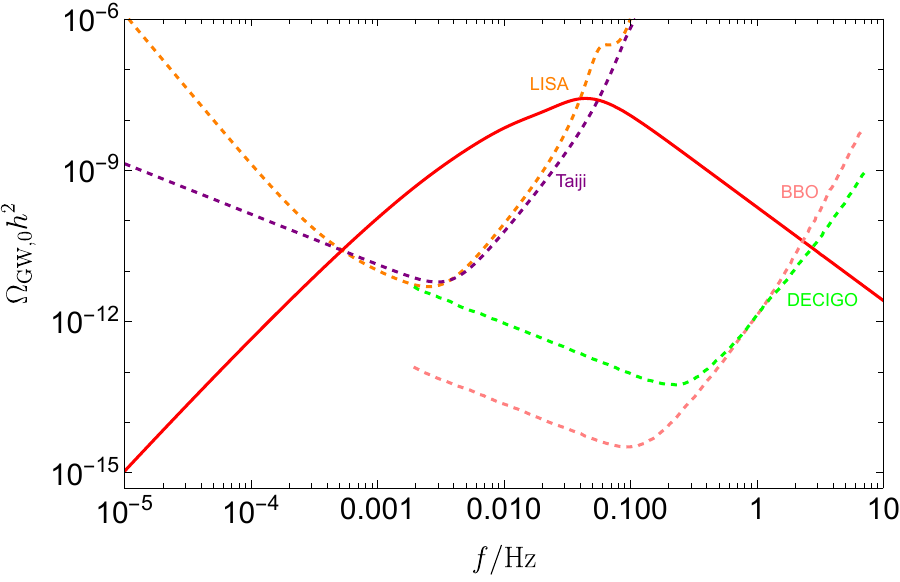}
    \caption{The current energy spectrum of GW as a function of frequency. The red solid curve represents the predicted spectrum of SIGW. The dashed lines depict the expected sensitivity curves for the proposed space-based interferometers: LISA (orange), Taiji (purple), DECIGO (green), and BBO (pink), which are summarized in~\cite{Moore:2014lga}.}\label{quintic_GW}
\end{figure*}

In the radiation-dominated era, the enhanced curvature perturbations responsible for generating a substantial PBH abundance inevitably source a significant SIGW through second-order cosmological perturbations. These SIGWs are predominantly produced when the corresponding scalar modes re-enter the horizon, and their growth saturates shortly thereafter as the scalar perturbations decay. We omit the detailed derivation and directly present the formula for energy spectrum of GWs, which is given by~\cite{Kohri:2018awv,Domenech:2021ztg}
\begin{align}\label{IGW}
		\Omega_{\mathrm{GW}}&(\eta,k) = \frac{1}{12} \int^\infty_0 dv \int^{|1+v|}_{|1-v|}du \left( \frac{4v^2-(1+v^2-u^2)^2}{4uv}\right)^2\mathcal{P}_\mathcal{R}(ku)\mathcal{P}_\mathcal{R}(kv)\nonumber\\
		&\left( \frac{3}{4u^3v^3}\right)^2 (u^2+v^2-3)^2\nonumber\\
		&\left\{\left[-4uv+(u^2+v^2-3) \ln\left| \frac{3-(u+v)^2}{3-(u-v)^2}\right| \right]^2  + \pi^2(u^2+v^2-3)^2\Theta(v+u-\sqrt{3})\right\}.
	\end{align}
Taking the thermal history of the Universe into consideration, the current energy spectrum of GWs can be expressed as
	\begin{equation}
		\Omega_{\mathrm GW,0}(k)= \Omega_{\gamma,0} \left(\frac{g_{\star,0}}{g_{\star,\mathrm eq}}\right)^{1/3}  \Omega_{\mathrm GW}(\eta_{\mathrm eq},k),
	\end{equation}
	where $\Omega_{\gamma,0}$ denotes the current density parameter of radiation, while $g_{\star,0}$ and $g_{\star,\mathrm eq}$ represent the effective relativistic degrees of freedom at the present time and at the epoch of radiation–matter equality $\eta_{\mathrm eq}$, respectively. Based on our parameter choice, the peak of the  energy spectrum can be estimated as $\Omega_{\mathrm GW, p}\sim\mathcal{O}(10^{-5}\mathcal{P}_\mathcal{R}^2)$. For a primordial curvature perturbation power spectrum $\mathcal{P}_\mathcal{R}\sim\mathcal{O}(0.1)$, this yields $10^{-7}$, in agreement with the amplitude shown in Fig.~\ref{quintic_GW}. As illustrated, the corresponding SIGW spectrum falls within the sensitivity bands of LISA~\cite{LISA:2017pwj} and Taiji~\cite{Ruan:2018tsw}, while remaining above the projected sensitivity curves of BBO and DECIGO~\cite{Kawamura:2011zz} across much of their frequency ranges. These features position the signal as an observable target for next-generation space-based interferometers.

\section{Summary and discussion}\label{IV}
In this paper, we have investigated the viability of polynomial potential inflation in light of the latest ACT data. We focused on the  cases of $n=2$, $3$, $4$ and $5$.  We have confronted their predictions with updated observational constraints on the scalar spectral index $n_s$, tensor-to-scalar ratio $r$, running of the spectral index $\alpha_s$ and running of the running of the spectral index $\beta_s$.

{For the quadratic potential ($n=2$), we find that the model can accommodate the ACT data at $1\sigma$ level within a limited range of the dimensionless coefficients $a_1$ and $a_2$. However, when imposing the additional constraint on the number of $e$-folds $N$, the model struggles to accommodate the ACT data within the standard e-folding number range of $N \approx 50$--$60$, as shown in Fig.~1. While larger values (e.g., $N=70$) can improve consistency, they are less typical based on thermal history considerations, highlighting the model's limitations.}

The cubic ($n=3$) and quartic potentials ($n=4$)  introduce additional parameters, providing greater flexibility in fitting observational data. By including the running $\alpha_s$  
and, for $n=4$, the running of the running $\beta_s$, we show that these models can achieve a good fit to ACT data for a range of $N$ values, including $N=60$. 

The introduction of a quintic term ($n=5$) enables significant phenomenological expansion. While lower-order models adequately describe CMB-scale observations, the quintic potential generates an approximate inflection point that triggers an ultra-slow-roll phase . This mechanism produces a substantial amplification of curvature perturbations on small scales, essential for PBH formation. Remarkably, the parameter set achieving CMB compatibility ($n_s=0.9764\,, r=4.8\times10^{-7}$) simultaneously generates a power spectrum peak at $k\approx 10^{13} {\mathrm Mpc^{-1}}$, demonstrating the model's capacity to bridge large and small-scale physics. Although reconciling the observational constraints from ACT with  PBH formation may appear to require the fine-tuning of multiple parameters, this burden is greatly alleviated by introducing a physical cutoff scale, $\Lambda$. Actually, the inflaton field can be rescaled to normalize the parameter $a_4 $ to unity. Consequently, the viability of this model depends primarily on the fine-tuning of a single remaining parameter, $a_5$, typically to a precision of order $\delta a_5 \sim 10^{-8} $~\cite{Hertzberg:2017dkh,Cole:2023wyx}. In the present paper, we focus solely on the phenomenology of polynomial inflation, without addressing the applicability of the theory at high energy scales.

{For $n = 2$ to $n = 4$, our analytical results are exact within the slow-roll approximation with $\phi_* = 0$. For $n = 5$, our treatment relies on the small-field approximation valid during the first slow-roll phase ($\phi < \phi_1 \ll 1$), which is appropriate for interpreting large-scale observations such as the ACT data. The analysis of small-scale dynamics near the inflection point, relevant for PBH production, would require including higher-order terms in the potential.}
Our results highlight the importance of revisiting and refining inflationary models in light of the latest observational data. The polynomial potential, despite its simplicity, can still provide a viable framework for explaining the early Universe, particularly when extended to higher-order terms. However, the model's parameter space is increasingly constrained by the precise measurements of the CMB anisotropies. To further test the polynomial potential inflation and other inflationary models, future CMB experiments with improved sensitivity and resolution, such as CMB-S4~\cite{CMB-S4:2016ple}, Simons Observatory~\cite{SimonsObservatory:2018koc}, and LiteBIRD~\cite{LiteBIRD:2022cnt}, will be crucial. These experiments are expected to provide even tighter constraints on $n_s$, $r$, and $\alpha_s$, allowing for a more stringent model selection and parameter estimation.

In conclusion, polynomial potential inflation remains a compelling framework for describing early universe dynamics, with higher-order terms ($n\ge4$) showing particular promise in reconciling ACT observations with the generation of small-scale structure. The interplay between CMB measurements and GW astronomy presents exciting opportunities for further refining our understanding of inflationary physics.


\begin{acknowledgments}
{We thank the anonymous referee for the valuable and constructive suggestions.}
This work has been supported by the National Key Research and Development Program of China (No. 2023YFC2206704). ZZP is supported by  the National Natural Science Foundation of China under Grant No. 12447166.
ZCC is supported by the National Natural Science Foundation of China under Grant No.~12405056, the Natural Science Foundation of Hunan Province under Grant No.~2025JJ40006, and the Innovative Research Group of Hunan Province under Grant No.~2024JJ1006. 
LL is supported by the National Natural Science Foundation of China (Grant No.~12505054, 12447101 and ~12433001) and the Fundamental Research Funds for the Central Universities.
\end{acknowledgments}

\bibliographystyle{apsrev4-1}
\bibliography{ref}

\end{document}